\documentclass[preprint,twocolumn]{jpsj3}
\usepackage{txfonts}
\usepackage{amsmath}
\usepackage{amssymb}%
\usepackage[dvipdfmx]{graphicx,color}
\usepackage{dcolumn}
\usepackage{bm}
\usepackage{ulem,color}

\newcommand{\lag}{\langle}
\newcommand{\rag}{\rangle}
\newcommand{\Lt}{\left}
\newcommand{\Rt}{\right}

\newcommand{\hspc}{\hspace{1em}}

\newcommand{\mrm}{\mathrm}
\newcommand{\mcl}{\mathcal}

\title{Phenomenological Magnetic Model in Tsai-Type Approximants}
\author{Takanori Sugimoto$^1$\thanks{sugimoto.takanori@rs.tus.ac.jp}, Takami Tohyama$^1$, Takanobu Hiroto$^2$, and Ryuji Tamura$^3$}
\inst{$^1$Dept. Appl. Phys., Tokyo Univ. Sci., Katsushika, Tokyo 125-8585, Japan\\
$^2$ Dept. Adv. Mater. Sci., Univ. Tokyo, Kashiwa, Chiba 277-8561, Japan\\
$^3$Dept. Mat. Sci. Tech., Tokyo Univ. Sci., Katsushika, Tokyo 125-8585, Japan} 

\abst{
Motivated by recent discovery of canted ferromagnetism in Tsai-type approximants Au-Si-RE (RE = Tb, Dy, Ho), we propose a phenomenological magnetic model reproducing their magnetic structure and thermodynamic quantities. In the model, cubic symmetry ($m\bar{3}$) of the approximately-regular icosahedra plays a key role in the peculiar magnetic structure determined by a neutron diffraction experiment. Our magnetic model does not only explain magnetic behaviors in the quasicrystal approximants, but also provides a good starting point for the possibility of coexistence between magnetic long-range order and aperiodicity in quasicrystals.
}

\kword{Quasicrystal, Ferromagnetism, Tsai cluster}

\begin{document}
\maketitle

Since the astonishing discovery of quasicrystals (QCs) in 1982,~\cite{shechtman84,goldman93} a question whether a long-range ordered state is allowed in aperiodic systems or not has been intensively investigated for three decades.
However, several preceding studies on magnetic properties at low temperatures in QCs have shown the difficulty of obtaining the long-range ordered state,~\cite{sato98,islam98} although there has been the theoretical prediction that a simple antiferromagnetically-ordered state is allowed in body-centered icosahedral QCs.~\cite{lifshitz98} 
Since the absence of long-range order has estranged QCs from a variety of applications, ordered states are long-awaited to be obtained in QCs.~\cite{goldman14}
 
\begin{figure*}[htb]
\centering
\includegraphics[width=0.8\textwidth]{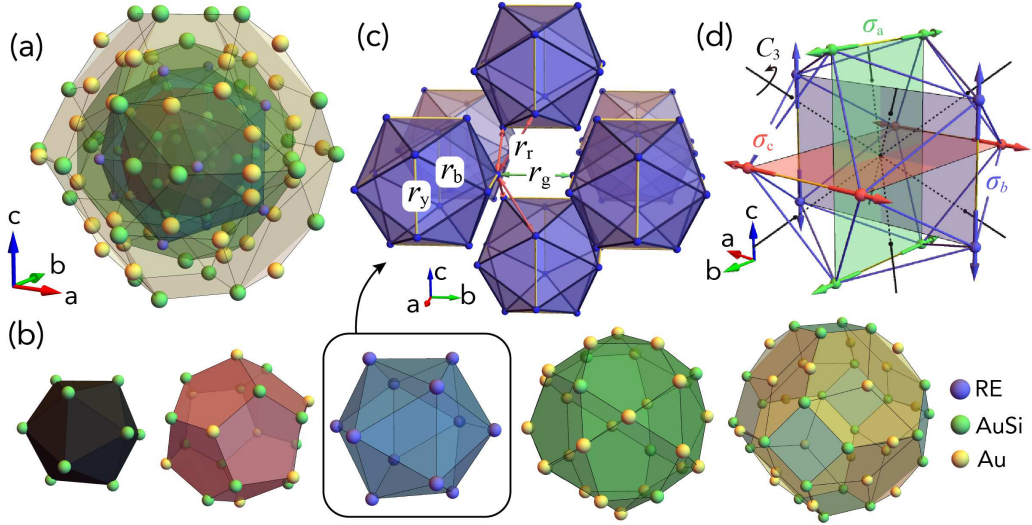}
\caption{
(Color online) Tsai-type cluster and crystal structure of Au-Si-RE. (a) Tsai-type cluster constructed by five polyhedral shells.~\cite{tsai97} (b) The polyhedra from left to right are arranged from the inside to the outside, respectively. Among the polyhedra, only icosahedral vertices have RE ions. In Tsai-type CQs and approximants, the Tsai-type cluster shares a hexagonal surface of the outermost polyhedral shell (a distorted rhombic triacontahedron) with a neighboring cluster. (c) Body-centered crystal of RE icosahedra. Ten-neighboring sites exist around a RE site, and the neighboring bonds are classified into four types according to length: $r_{\mrm{b}}$=5.50~\AA, $r_{\mrm{y}}$=5.42~\AA, $r_{\mrm{r}}$=5.42~\AA, and $r_{\mrm{g}}$=5.72~\AA for Au-Si-Tb, $r_{\mrm{b}}$=5.51~\AA, $r_{\mrm{y}}$=5.43~\AA, $r_{\mrm{r}}$=5.43~ \AA, and $r_{\mrm{g}}$=5.73~\AA for Au-Si-Dy, $r_{\mrm{b}}$=5.51~\AA, $r_{\mrm{y}}$=5.43~\AA, $r_{\mrm{r}}$=5.42~\AA, and $r_{\mrm{g}}$=5.73~\AA for Au-Si-Ho. (d) The SIA vectors and crystalline symmetries in an icosahedron. There are three inversion ($\mcl{\sigma}_a$, $\mcl{\sigma}_b$ and $\mcl{\sigma}_c$) and four three-fold rotational symmetry ($\mcl{C}_3$) operators. The SIA vectors denoted by arrows on vertices are obtained by comparing theoretical spin structure with experimental one and by applying the rule 2 mentioned in the main text.}
\label{fig1}
\end{figure*}

Besides QCs, resent works have shown that some icosahedral approximants exhibit superconductivity~\cite{deguchi15} and (anti-)ferromagnetism.~\cite{tamura10,hiroto14}
The approximants contain the Tsai-type~\cite{tsai94} cluster (Fig.~\ref{fig1}(a)) composed of several polyhedral shells (Fig.~\ref{fig1}(b)) including an icosahedral shell. 
The clusters form body-centered cubic (bcc) structure (Fig.~\ref{fig1}(c)) by sharing a hexagonal surface of the outermost polyhedral shell (a distorted rhombic triacontahedron). 
As a result, the clusters in the approximants are just slightly distorted in contrast to those in the Tsai-type QCs. 
Since the approximants can be treated with conventional analysis in periodic crystals but have the same local structure as QCs, the discovery of the long-range order mechanism will indicate the possibility of ordered states in QCs. 
Clarifying the mechanism of ferromagnetism is thus a starting point of solving the long-standing problem in QCs.

Among icosahedral approximants, canted ferromagnetism has recently been reported in Au-Si-REs (RE = Tb, Dy, Ho),~\cite{hiroto14} which have the same space group as Au-Si-Gd exhibiting simple ferromagnetism.~\cite{hiroto13}
The compound Au-Si-RE and Au-Si-Gd have locally the same structure as the Tsai-type QCs,~\cite{Gebresenbut13} where magnetic moment is represented as the total angular momentum of localized f-electrons with a strong spin-orbit coupling in rare-earth ions located at vertices of icosahedra (see Fig.~\ref{fig1}(b)).
The RE ion has angular momentum in f-electron orbit but the Gd ion does not.
This difference should explain the presence of the canted ferromagnetism in Au-Si-RE.

In this Letter, we examine magnetic behaviors in icosahedral approximants Au-Si-RE. 
Taking into account strong spin-orbit coupling coming from finite angular momentum in RE and the Ruderman-Kittel-Kasuya-Yosida (RKKY) interaction,~\cite{ruderman54,kasuya56,yosida57} we construct a phenomenological magnetic model Hamiltonian.
Their parameter values are determined by reproducing the magnetic susceptibility and the magnetization curve.
It is also found that cubic ($m\bar{3}$) symmetry of icosahedral clusters in bcc structure determines the direction of single-ion anisotropy (SIA).
Putting these values into our model, we obtain magnetic structure consistent with recent neutron diffraction experiment.~\cite{hirotoTBP}
Our magnetic model that explains magnetic behaviors in the Tsai-type approximants will give a hint for the magnetic order in QCs.

To discuss the magnetic behaviors in Au-Si-RE, we consider three-types of magnetic interactions in a classical vector spin model: the SIA term caused by spin-orbit coupling, the RKKY interaction induced by the s-f coupling, and the Zeeman term. 
A resulting Hamiltonian reads
\begin{equation}
\mcl{H}= -D\sum_i \Lt(\hat{\bm{d}}_i\cdot \bm{S}_i\Rt)^2 - \sum_{\lag i,j\rag} J_{\lag i,j \rag} \bm{S}_i\cdot \bm{S}_j - \mu_{\mrm{eff}} H\sum_i \hat{\bm{h}}\cdot \bm{S}_i, \label{eq:ham}
\end{equation}
where $\bm{S}_i$ is the unit-length vector of spin with SO(3) on the $i$-th RE site, $D$ is the SIA energy of the spin projected to the direction of SIA with the unit vector $\hat{\bm{d}}_i$, $J_{\lag i,j \rag}$ is the RKKY exchange energy between neighboring RE sites $\bm{r}_i$ and $\bm{r}_j$, $\mu_{\mrm{eff}}$ is the effective magnetic moment of the spin, and $H$ is the magnitude of the magnetic field applied to the direction $\hat{\bm{h}}$ ($|\hat{\bm{h}}|=1$).

Assuming the isotropic Fermi wavenumber $k_F$ under the free Fermi gas approximation and the bond length $|\bm{r}_i-\bm{r}_j|$ estimated by crystal structure, the RKKY exchange energy may be given by $J_{\lag i,j \rag} = J f(2k_F |\bm{r}_i-\bm{r}_j|)$ with $J>0$ and $f(x)=(-x\cos x +\sin x)/x^4$.
In Au-Si-RE, four lengths of ten neighboring sites around a RE site, $r_{\mrm{b}}$, $r_{\mrm{y}}$, $r_{\mrm{r}}$, and $r_{\mrm{g}}$, shown in Fig.~\ref{fig1}(c), are similar and are expected to predominantly contribute to the RKKY interaction. The corresponding energies are denoted by $J_\alpha=Jf(2k_F r_\alpha)$ with $\alpha=\mrm{b,y,r,g}$ hereafter. The value of $k_F$ can be estimated from the Fermi gas approximation, $k_F=(3\pi^2 n)^{1/3}$ with the density of free electrons $n=0.100\> \AA^{-3}$ (Au-Si-Tb), $0.098\> \AA^{-3}$ (Au-Si-Dy), and $0.098\> \AA^{-3}$ (Au-Si-Ho), respectively.~\cite{ishikawa16}
Taking into account the lattice constant for RE= Tb, Dy, Ho,~\cite{hiroto14} $k_F$ is obtained as shown in Table~\ref{tab1}. 

In order to describe magnetic properties in Au-Si-RE by (\ref{eq:ham}), we have to determine RE-dependent parameter values of $D$, $J$, and $\mu_{\mrm{eff}}$, together with the direction of $\hat{\bm{d}}_i$. We will describe the procedure to determine them below.

To introduce the cubic symmetry of the approximants into our model, we assume that the SIA unit vector $\hat{\bm{d}}_i$ satisfies following rules based on the three-fold rotational, inversion, and also translational symmetries ($\mcl{C}_3$, $\mcl{\sigma}$, and $\mcl{K}$) in the crystalline point group $m\bar{3}$, except for the five-fold rotational symmetry in a regular icosahedron.
\begin{itemize}
\item[] \hspace{-5mm} {\it Rule}~1. The translational symmetry, which projects a RE site onto an equivalent site in another icosahedron, keeps the SIA vectors invariant: $\mcl{T}: \bm{r}_i \to \bm{r}_j \Rightarrow \hat{\bm{d}}_i=\hat{\bm{d}}_j$.
\item[] \hspace{-5mm} {\it Rule}~2. In an icosahedron, the SIA vector follows the three-fold rotational and inversion symmetries: $\mcl{C}_3:\bm{r}_i \to \bm{r}_j \Rightarrow \hat{\bm{d}}_j=\mcl{C}_3(\hat{\bm{d}}_i)$ and $\mcl{\sigma}:\bm{r}_i \to \bm{r}_j \Rightarrow \hat{\bm{d}}_j=\mcl{\sigma}(\hat{\bm{d}}_i)$.~\cite{note1}
\end{itemize}
The rule 2 is natural since the local spin anisotropy would be determined by surrounding crystal field.
By the rules, three SIA vectors and their inversion are allowed.
An important feature of the SIA vectors is that their rotational degree of freedom is restricted in a rotational plane due to the inversion symmetry,~\cite{note2} that is, the SIA vector has a degree of freedom associated with a global SO(2) rotation in our model.
Suppose the direction of the anisotropy is locally determined at a certain site, then the others are automatically fixed by the rules based on crystalline symmetry operators $\mcl{C}_3$, $\mcl{\sigma}$, and $\mcl{T}$.
Figure~\ref{fig1}(d) shows an example of possible SIA vectors satisfying the rule 2 in an icosahedron with the crystalline symmetries $\mcl{C}_3$ and $\mcl{\sigma}$, where the SIA vectors are parallel to each other in the same plane, and perpendicular to each other in the other planes.
We will determine the SIA vectors in real Au-Si-RE by comparing theoretical spin structure with experimental one by the neutron diffraction.~\cite{hirotoTBP}
The example shown in Fig.~\ref{fig1}(d) is actually the case of real Au-Si-RE as discussed below.

The magnetic susceptibility in the effective Hamiltonian is given by,
\begin{equation}
\chi(\beta)=\frac{1}{\beta}\frac{\partial^2 \ln \mrm{Tr} [e^{-\beta \mcl{H}}]}{\partial H^2} \Bigg|_{H\to0}
\end{equation}
where $\mrm{Tr}[\cdots]$ represents an integral over the spin coordinates $\prod_i \int \mcl{D}\bm{S}_i \cdots$, and $\beta=1/k_BT$ is an inverse temperature with the Boltzmann factor $k_B$.
With this equation, we can obtain a high-temperature expansion of the magnetic susceptibility within the third order of $\beta$,
\begin{equation}
\chi(\beta)=\chi^{(1)}\beta+\frac{\chi^{(2)}}{2!}\beta^2+\frac{\chi^{(3)}}{3!}\beta^3+O(\beta^4) \label{eq:chia}
\end{equation}
with
\begin{align}
\chi^{(1)}&=\frac{N\mu_{\mrm{eff}}^2}{3}, \hspc \chi^{(2)}=\frac{2N\mu_{\mrm{eff}}^2}{9}(4J_{\mrm{b}}+J_{\mrm{y}}+4J_{\mrm{r}}+J_{\mrm{g}}),\notag \\
\chi^{(3)}&=\frac{4N\mu_{\mrm{eff}}^2}{9}[6J_{\mrm{b}}^2+6J_{\mrm{r}}^2+16J_{\mrm{b}}J_{\mrm{r}}+J_{\mrm{y}}J_{\mrm{g}}+4(J_{\mrm{b}}+J_{\mrm{r}})(J_{\mrm{y}}+J_{\mrm{g}})], \label{eq:chib}
\end{align}
where $N$ is the number of RE sites.
It is interesting to notice that the SIA term ($D$) does not appear in the magnetic susceptibility within the third order of $\beta$.
The disappearance of the SIA contribution originates from three orthogonal SIA axes preserving the cubic symmetry.
Consequently, the susceptibility in our model is independent from the direction of applied magnetic field at high temperatures.
It also concludes that there is no difference in the high-temperature susceptibilities of a single crystal and a polycrystal of Au-Si-RE.
Thus, the proposed model for Au-Si-RE is an interesting example, for which the high-temperature susceptibility does not depend on the axis of an applied field in spite of the presence of the easy axis of the magnetic moment.

\begin{figure}[htb]
\centering
\includegraphics[width=0.45\textwidth]{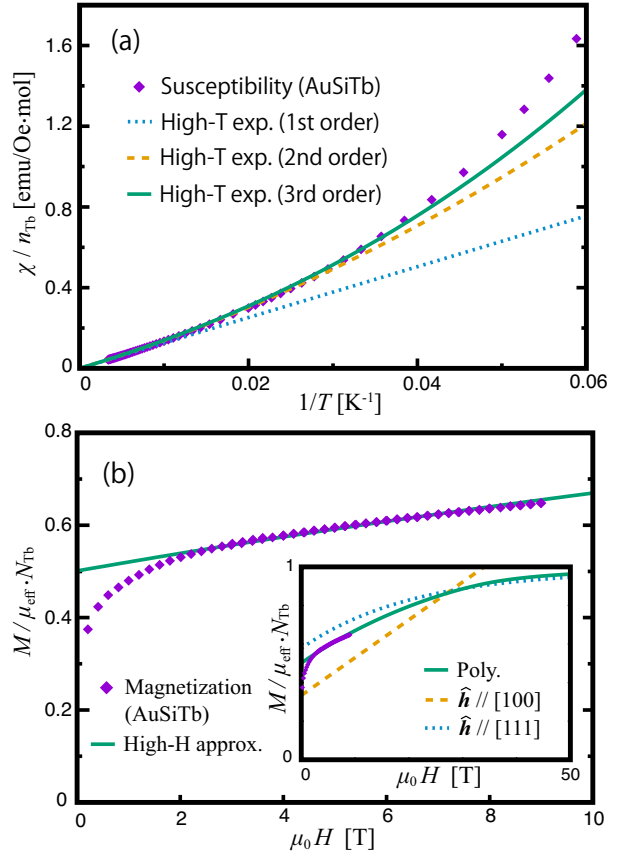}
\caption{(Color online) (a) The comparison of high-temperature-expansion susceptibility given by Eqs.~(\ref{eq:chia}) and (\ref{eq:chib}) with the experimental data (diamonds) of Au-Si-Tb.~\cite{hiroto14} The fitting to the experimental data up to the third order of $\beta$ is shown. The values of $J$ and $\mu_{\mrm{eff}}$ listed in Table~\ref{tab1} are obtained from the fitting.
(b) The comparison of calculated high-field magnetization curve (sold line) with  experimental data (diamonds) of Au-Si-Tb.~\cite{hiroto14} The SIA energy $D$ is estimated by comparing of the model and the experiment at high magnetic fields. The inset represents the magnetization curves with magnetic fields $\hat{\bm{h}}\parallel [100]$ and $\hat{\bm{h}}\parallel [111]$, as compared with the polycrystalline case.}
\label{fig2}
\end{figure}

Since high-temperature magnetic susceptibility depends on the RKKY interactions and the effective magnetic moment, we can estimate the magnitude of $J$ and $\mu_{\mrm{eff}}$ by fitting Eq.~(\ref{eq:chia}) to the experimental data of Au-Si-RE.~\cite{hiroto14} 
Figure~\ref{fig2}(a) shows the experimental data and fitted results for RE=Tb, where the fitting is performed with increasing the order of $\beta$ by using Eq.~(\ref{eq:chib}).
The evaluated value of $J$ for Au-Si-Tb is 12,000~K, resulting in $J_\mrm{b}=3.0$~K, $J_\mrm{y}=3.2$~K, $J_\mrm{r}=3.2$~K, and $J_\mrm{g}=1.9$~K ,i.e., $1.9 \mrm{K} \le J_\alpha \le 3.2 \mrm{K}$ ($\alpha=\mrm{b,y,r,g}$) as shown in Table~\ref{tab1}.
The value of the effective magnetic moment is $\mu_{\mrm{eff}}=10.1\mu_B$ for Au-Si-Tb.
We also obtain the values of $J$ ($J_\alpha$) and $\mu_{\mrm{eff}}$ for Au-Si-Dy and for Au-Si-Ho (see Table~\ref{tab1}), by fitting the experimental susceptibility data with Eq.~(\ref{eq:chia}) (not shown).

It is interesting to compare $\mu_{\mrm{eff}}$ estimated from experimental data with the effective moment $\mu_{\mrm{eff}}^{\mrm{SI}}$ in an isolated single ion of RE, the latter of which is given by $\mu_{\mrm{eff}}^{\mrm{SI}}=\mu_Bg_L\sqrt{L_{\mrm{tot}}(L_{\mrm{tot}}+1)}\>$, where the Land\'e $g$ factor $g_L=1+[L_{\mrm{tot}}(L_{\mrm{tot}}+1)+L_{\mrm{spin}}(L_{\mrm{spin}}+1)-L_{\mrm{orb}}(L_{\mrm{orb}}+1)]/[2L_{\mrm{tot}}(L_{\mrm{tot}}+1)]$ with the orbital, spin, and total angular momenta, denoted by $L_{\mrm{orb}}$, $L_{\mrm{spin}}$, and $L_{\mrm{tot}}$, respectively.
We find that $\mu_{\mrm{eff}}$ is equal to $\mu_{\mrm{eff}}^{\mrm{SI}}$ within an error of 5 percent as seen in Table~\ref{tab1}.
This implies that single-ion character in RE is strongly preserved in Au-Si-RE as expected.

\begin{table*}[htb]
  \centering
  \caption{Estimated parameters in the effective model. The effective magnetic moment $\mu_{\mrm{eff}}$ and the RKKY constant $J$ are obtained by comparing the calculated susceptibility with experimental ones for Au-Si-RE (RE = Tb, Dy, Ho) at high temperatures. For comparison, the magnetic moment $\mu_{\mrm{eff}}^{\mrm{SI}}$ from an isolated RE ion is also listed. The effective RKKY energies $J_\alpha$ ($\alpha=\mrm{b,y,r,g}$) between four-types of neighboring bonds are obtained by the RKKY constant $J$, the bond lengths $r_\alpha$, and the Fermi wavenumber $k_F$ for each materials and the minimum and maximum value among $J_\alpha$ is listed. The SIA energy $D$ is estimated by comparing calculated magnetization curves with experimental ones.}
  \label{tab1}
  \begin{tabular}{ccccccc}
    \hline
    Compound & $\mu_{\mrm{eff}}$ [$\mu_B$] & $\mu_{\mrm{eff}}^{\mrm{SI}}$ [$\mu_B$] & $k_F$ [$\AA^{-1}$] & RKKY constant $J$ [K] & RKKY energies $J_\alpha$ [K] & SIA energy $D$ [K] \\
    \hline
    Au-Si-Tb & 10.1 & 9.7 & 1.44 & 12,000 & 1.9 - 3.2 & 120\\
    Au-Si-Dy & 11.1 & 10.6 & 1.43 & 6,000 & 1.0 - 1.6 & 130\\
    Au-Si-Ho & 10.6 & 10.6 & 1.43 & 3,500 & 0.6 - 0.9 & 60\\
    \hline
  \end{tabular}
\end{table*}

As shown in Fig.~\ref{fig2}(b), the experimental magnetization curves of Au-Si-RE~\cite{hiroto14} do not saturate up to 9~T, which is larger than the estimated maximum RKKY exchange energy (3.2~K ($\cong$ 0.47~$\mu_{\mrm{eff}}k_B^{-1}$ T) as shown in Table~\ref{tab1}).
This implies that the high-field magnetization is not controlled by the RKKY interaction but another energy scale coming from the SIA energy.
Hence, we can estimate the SIA energy $D$ from high-field magnetization curves by assuming that the ground state at high magnetic fields can be described by the SIA and Zeeman terms.
In this assumption, an angle between a local spin and the direction of the magnetic field, $\theta_i=\cos^{-1} (\hat{\bm{h}}\cdot\bm{S}_i)$, which is dependent on the strength of the magnetic field $H$, is obtained as a solution of a following variational equation,
\begin{equation}
\frac{\partial \mcl{H}}{\partial \theta_i}\cong -D\sin[2(\alpha_i-\theta_i)]+\mu_{\mrm{eff}}H\sin\theta_i=0, \label{eq:mh}
\end{equation}
where $\alpha_i=\cos^{-1}(\hat{\bm{h}}\cdot\hat{\bm{d}}_i)$.

With a solution of spin angles as a function of the magnetic field $\theta_i(H)$, the magnetization curve is obtained as a sum of spin inner-product components, $M(H)=\mu_{\mrm{eff}}\sum_i \cos\theta_i(H)$.
The magnetization curve of a single crystal is anisotropic with respect to the orientation of the magnetic field, {\it e.g.}, the magnetization curves with magnetic fields parallel to [100] direction and the case of [111] direction shown in the inset of Fig.~\ref{fig2} (b).
To compare the mganetization curve measured for polycrystalline samples of Au-Si-RE, we numerically integrate the magnetic moments $M(H)$ over whole direction of $\hat{\bm{d}}$, where 10,000 $\hat{\bm{d}}$ samples are taken into account keeping the icosahedral structure, at a fixed magnetic field.
We fit the integrated magnetization curves on the experimental data at high magnetic fields (see Fig.~\ref{fig2} (b)).
Since the integrated magnetization curve is isotropic under the rotation of the magnetic field: $\mcl{R}: \hat{\bm{h}}\to\hat{\bm{h}}^\prime $, the direction of the SIA vectors cannot be determined by the comparison.
However, we can find a big anisotropy in the magnetization curves of the single crystal, and thus propose how to clarify the direction of the SIA vectors by using the magnetization curves.

Table~\ref{tab1} shows estimated SIA energies in our model obtained from high-field magnetization.
As expected, the SIA energies are much larger than the RKKY energies in Au-Si-RE and thus are dominant contribution in our magnetic model. 
In contrast to Au-Si-RE, such a large SIA energy cannot be expected in Au-Si-Gd, since Gd ion exhibits zero orbital angular momentum.
This would be the origin of the difference of magnetic behaviors between Au-Si-RE~\cite{hiroto14} and Au-Si-Gd,~\cite{hiroto13} the latter of which exhibits simple ferromagnetism. 
Therefore, the cubic symmetry of the SIA vectors in the RE icosahedra, rather than the slight difference in the distance between RE sites in the icosahedron, plays a key role to generate the magnetic structure.
 
The spin structure in Au-Si-Tb at a low temperature has been reported by using a neutron diffraction experiment.~\cite{hirotoTBP}
The experiment has shown {\it orthogonal} ferromagnetism on three sublattices in Tb sites.
In order to clarify whether such three-sublattice ferromagnetism is possible or not in our magnetic model, we investigate the spin configuration in the ground state by using the simulated-annealing (SA) method, through which the direction of SIA is determined.
The SA calculation with $10^6$ steps of the annealing is done for the $10\times10\times10$ unit cells, including 2 icosahedral clusters per cell, under periodic boundary conditions.
We confirm that the ground-state energy per site converges to $122.93$~K, which is close to the sum of the SIA energy $D$ and the intra-plane effective RKKY energy per site $(J_{\mrm{y}}+J_{\mrm{g}})/2$.

\begin{figure}[htb]
\centering
\includegraphics[width=0.35\textwidth]{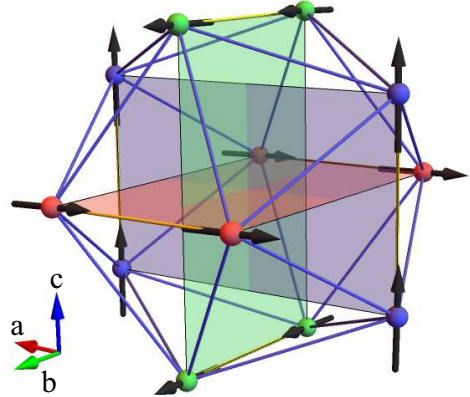}
\caption{(Color online) Calculated spin structure on an icosahedral cluster obtained by the SA simulation for $10\times10\times10$ unit cells. The black arrows represent the spin moment averaged over the equivalent 1000 sites. There are ferromagnetic spin structures on a-b, b-c, and c-a planes. This orthogonal ferromagnetism forming three-sublattice structure is consistent with the neutron diffraction experiment within an experimental error.~\cite{hirotoTBP} The SIA vectors are fixed by using the global rotational degree of freedom, corresponding to the experimental result of the neutron diffraction.~\cite{hirotoTBP} The average of skewed angles $\lag \cos^{-1} \hat{\bm{d}}_i\cdot\bm{S}_i\rag = 0.054 \pm 0.014$ is also consistent with the experiment.}
\label{fig3}
\end{figure}

Figure~\ref{fig3} shows a resulting ground-state spin structure on an icosahedral cluster, where the spin moment on Tb ion denoted by the magnitude of the arrows is obtained by averaging over the equivalent 1000 sites and its averaged value equals $\mu_{\mrm{eff}}$ within an error of $3\times 10^{-3}$ percent.
We can find orthogonal ferromagnetism on three sublattices, where four spins on the same plane arranged nearly ferromagnetically.
This is consistent with the neutron diffraction experiment.~\cite{hirotoTBP}
Such nearly in-plane ferromagnetism in our model is supported by a small skew angle of spin, $\lag \cos^{-1} \hat{\bm{d}}_i\cdot\bm{S}_i\rag = 0.054 \pm 0.014$, which is roughly given by $\sin^{-1} (4\lag J \rag/D) \sim 0.08$ with the inter-plane effective RKKY energy per site $\lag J \rag = (J_{\mrm{b}}+J_{\mrm{r}})/2$.
In our model, the three-sublattice structure is caused by the original crystal structure in Au-Si-RE, where the five-fold rotational symmetry is broken in the space of the SIA vectors in icosahedral RE clusters.
Since the proposed Hamiltonian can reproduce various magnetic properties, it would be a minimal model to describe the physics of magnetism in the Tsai-type rare-earth approximants.

In summary, we have investigated a phenomenological magnetic model to reproduce the experimental results in Au-Si-RE (RE = Tb, Dy, Ho).
The model is based on a large spin-orbit coupling and the RKKY interactions between localized spins at the rare-earth sites.
The magnetic susceptibility and magnetization curves are used to estimate the parameters in our model.
The estimated energies of the single-ion anisotropy are much larger than estimated RKKY interactions.
We confirm that the dominant single-ion anisotropy induces the orthogonal ferromagnetism on three-sublattice structure at low temperatures by employing the simulated-annealing method.~\cite{note3}
Since the three-sublattice structure originates from the crystal structure of the approximately-regular icosahedra, the orthogonal ferromagnetism is common to the Tsai-type rare-earth approximants, where the single-ion-anisotropy vectors in the rare-earth icosahedra break the five-fold rotational symmetry.
If the five-fold rotational symmetry is kept, which may correspond to the situation of the Tsai-type quasicrystal, the axes of the single-ion anisotropy are grouped into 6 types at least.
In the real quasicrystals, the crystal fields have also several variations reflecting on the local structures, which should group the anisotropy into more various categories.
For this reason, the magnetism in quasicrystals may have a locally-freezing moment along the anisotropy categorized by the local symmetries, which is expected to exhibit a glass-like behaviors.

\begin{acknowledgment}
This work was supported by Toyko University of Science.
Numerical computation in this work was carried out on the supercomputers at the Supercomputer Center at Institute for Solid State Physics, University of Tokyo.
\end{acknowledgment}



\begin{thebibliography}{9}
\bibitem{note1} Since the spin vectors and magentic fields are puedo-vectors, the parity projection them invariant. On the other hand, the SIA vectors originates from the crystal fields around rare-eearth sites, which shoud be odd under the parity and inversion.
\bibitem{note2} Although the SIA vectors can break the crystalline symmetry spontaneously, the precceding neutron-diffraction study~\cite{hirotoTBP} does not indicate any obvious evidences associated with the symmetry breaking.
\end{thebibliography}
\end{document}